\documentclass{aa}
\usepackage{graphicx}
\usepackage{epsfig}
\usepackage[authoryear]{natbib}
\bibpunct{(}{)}{;}{a}{}{,} 

\begin{document}
\newcommand{\iras}{{{IRAS\,12556--7731}}}

\title{\iras: a  
``chamaeleonic'' lithium-rich M-giant \thanks {Based on 
HARPS observations collected at the La Silla Observations.}
\fnmsep\thanks{Figures \ref{NLi_07} and \ref{harps_TiO} 
are only available in electronic form.}
}
\author{ J.M.~Alcal\'a\inst{1}
  \and K.~Biazzo\inst{1} 
  \and E.~Covino\inst{1}
  \and A.~Frasca\inst{2}
  \and L.R.~Bedin\inst{3}
}

\offprints{J.M. Alcal\'a}
\mail{alcala@oacn.inaf.it}

\institute{
     INAF-Osservatorio Astronomico di Capodimonte, via Moiariello 16, I-80131 Napoli, Italy
\and INAF-Osservatorio Astrofisico di Catania, via S. Sofia 78, I-95123 Catania, Italy
\and Space Telescope Science Institute, 3800 San Martin Drive, Baltimore, MD 21218, USA
}

\date{Received ; accepted }

\abstract
{}
{In this letter we characterise \iras\ as the first lithium-rich M-type 
giant. Based on its late spectral type and high lithium content, and 
because of its proximity in angular distance to the Chamaeleon\,II 
star-forming region, the star was  misclassified as a young low-mass star 
in a previous work.}  
{Based on HARPS data, synthetic spectral modelling, and proper motions, 
we derive the astrophysical parameters and kinematics of the star and 
discuss its evolutionary status.} 
{This solar-mass red giant ($T_{\rm eff}=3460\pm60$\,K and $\log{g}=0.6\pm0.2$) 
is characterised by a relatively fast rotation ($v\sin{i}\sim8$\,km~s$^{-1}$),
slightly subsolar metallicity and a high-lithium abundance, $A$(Li)$=2.4\pm0.2$\,dex. 
We discuss \iras\ within the context of other known lithium-rich 
K-type giants. Because it is close to the tip of the red giant branch, \iras\ 
is the coolest lithium-rich giant known so far, and it is among the least 
massive and most luminous giants where enhancement of lithium has been 
detected. 
Among several possible explanations, we cannot preclude the possibility that 
the lithium enhancement and rapid rotation of the star were triggered by the 
engulfment of a brown dwarf or a planet.}
{}


\titlerunning{\iras: the so far coolest lithium-rich giant}
\authorrunning{Alcal\'a et al.}
\maketitle

\section{Introduction} 
\label{Sec:intro}

The source \object{\iras} was first identified by \citet{prusti92} 
in an IRAS survey of pre-main sequence (PMS) stars in the Chamaeleon\,II 
(Cha\,II) cloud. The source was associated with a relatively bright 
($R\approx$12\,mag) off-cloud star. A later analysis of the star's near-IR 
magnitudes led \citet{larson98} to conclude that its IR colours are more similar 
to those of giants than those of PMS stars. The source was then classified
 as a young stellar object candidate using selection criteria based on 
{\it Spitzer} mid-IR data \citep{alcala08}, but it lacks the IR-excess 
\citep{larson98, alcala08} typical of young stars. Since lithium is efficiently
destroyed by convective mixing in the interior of low-mass stars when 
the temperature at the bottom of the convective layer reaches about 
2.5$\times$10$^6$\,K, the presence of strong Li\,{\sc i} $\lambda$6707.8\,\AA\  
absorption represents an important criterion for identification of 
low-mass young stars. Thus, a late-type spectrum and the detection of 
a strong Li\,{\sc i} $\lambda$6707.8\,\AA\  absorption in two VLT-FLAMES/Giraffe 
spectra, have led \citet{spezzi08} to assign a young stellar object nature to \iras. 
Surprisingly, this star turned out to be over-luminous by about two orders of 
magnitude on the HR diagram with respect to other young stars in the Cha\,II 
cloud \citep{spezzi08}. This result and the conclusion by \citet{larson98} 
that \iras\ could be a giant star, triggered a new  high-resolution spectroscopic 
study. We characterise the star as a lithium-rich M-type giant. 
Several spectroscopic studies of K-type giants in the past two decades 
have provided evidence of lithium enhancements in the atmosphere of these 
stars \citep[][ and references therein]{kumar11}. 
We study \iras\ in the context of other well known lithium-rich 
giants and discuss possible mechanisms for its lithium enhancement.

\begin{table*}
\caption[ ]{\label{physpar} Physical parameters of \iras.}
\centering
\begin{tabular}{ccccccccc}  
\hline
 $T_{\rm eff}$ & $\log{g}$    & [Fe/H]   &  $v\sin{i}$    & $RV$ &   $A$(Li) &  $\log{(L/L_{\odot})}$ & $\log{(R/R_{\odot})}$ & $M/M_{\odot}$  \\ 
 (K)         &   (dex)       &  (dex)   &  (km s$^{-1}$) & (km s$^{-1}$)        &  (dex)   & 		      & 	     &  	      \\ 
\noalign{\smallskip}
\hline
\noalign{\smallskip}
 3460$\pm$60   &  0.6$\pm$0.2  & $-$0.08$\pm$0.20 &  8$\pm$3 & 67.68$\pm$0.02  &  $2.4\pm0.2$  &  2.99$\pm$0.15  &  1.9$\pm$0.2	  & 1.0$\pm$0.2 \\        
\noalign{\smallskip}
\hline

\end{tabular}
\tablefoot{$T_{\rm eff}$ and $\log{g}$: from the $\chi^2$ minimization;
 [Fe/H] and $v\sin{i}$: average values from $\chi^2$ minimization and MOOG results.}
\end{table*}

\begin{figure*}[ht] 
\vskip 3.3cm
\includegraphics{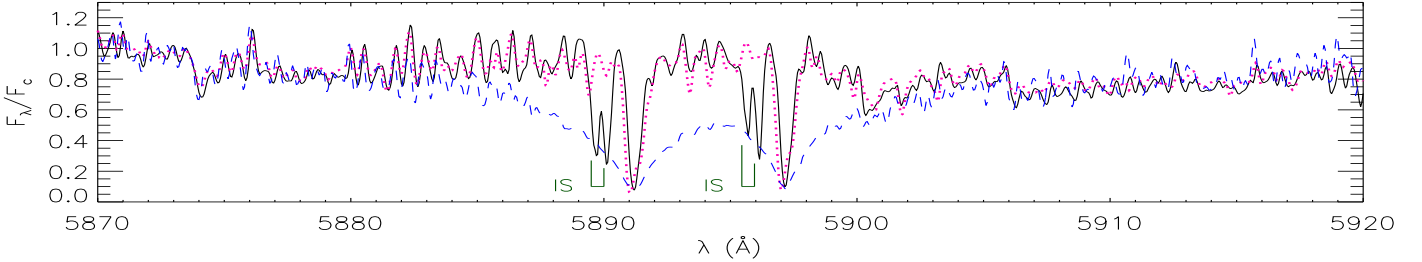}
 \caption{Portion of the HARPS spectrum of \iras\ in the \ion{Na}{i} D region. The M5\,III best template 
 (\object{HD\,175865}) is overplotted (dotted line). A dwarf M star with nearly the same temperature 
 (\object{HD\,173740}) is also overplotted with a dashed line. The interstellar absorption 
 components (IS) are also indicated.}
  \label{harps_spec}
\end{figure*}

\section{HARPS spectroscopy} 
\label{Sec:spectra}

With the aim of ascertaining the nature of \iras, 
two high-resolution (R$\sim$110,000) spectra (range from 3800\,\AA\ to 6900\,\AA), 
obtained in February 2-3, 2011 with the HARPS (High Accuracy Radial 
velocity Planet Searcher) spectrograph at the ESO 3.6\,m telescope in 
La Silla (Chile), were combined. 
The strong Li\,{\sc i} absorption line at $\lambda$6707.8\,\AA~was the 
first feature to be immediately confirmed, for which we measured an 
equivalent width of 550($\pm$10)\,m\AA. Likewise, the H$\alpha$ 
line is in absorption. Overall, a number of strong molecular absorption 
bands, mainly of titanium oxide, can be identified, and they confirm it
as a very cool object. 
The interstellar (IS) Na\,{\sc i}~D ($\lambda\lambda$5890,5996\,\AA) 
absorption components, distinguishable from the photospheric ones, 
are clearly detected (see Fig.~\ref{harps_spec}). The narrower \ion{Na}{i}\,D 
wings of \iras\ with respect to \object{HD\,173740} (see Fig.~\ref{harps_spec})
definitely exclude it as a main sequence star. 
 
The mean radial velocity of \iras, as drawn from the HARPS spectra, 
is $RV$=$67.68\pm$0.02\,km\,s$^{-1}$, which is far beyond the range 
of the Chamaeleon region \citep[$\sim$15$\pm2$\,km\,s$^{-1}$;][]{covino97}. 
Likewise, the proper motion components of the star, 
$\mu_{\alpha}\cos\delta$$=-13.1\pm$5\,mas~yr$^{-1}$, 
$\mu_\delta$$=10.7\pm$5\,mas~yr$^{-1}$,  as retrieved from the PPMXL Catalog 
\citep{roeser10}, are  not compatible with those of the Chamaeleon young 
stars \citep[e.g.,][]{frink98}. Therefore, we conclude  that the kinematics 
of \iras\ is inconsistent with those of young stars in the Chamaeleon region 
and that the star is unrelated to the star-forming cloud. 
The detection of the interstellar Na\,{\sc i} absorption components, 
for which we measure an $RV_{\rm IS}\sim 14\pm2$\,km\,s$^{-1}$  consistent with 
the radial velocity of Chamaeleon, is an indication that such interstellar 
components can be produced by the Cha\,II cloud itself and that \iras\ must 
be located at a much greater distance. Because it is unrelated to the star-forming 
region, \iras\ is most likely a field giant star, far behind the Cha\,II cloud. 
This confirms the suggestion by \citet{larson98}.

\section{Physical parameters and lithium abundance} 
\label{Sec:param}
 
Using the combined (S/N$\approx$30) HARPS spectrum, we obtained 
first estimates of the physical parameters by a  procedure outlined 
in detail in Appendix~\ref{appendix1}. This procedure provides us with 
the best effective temperature, gravity, and metallicity by comparing 
the combined HARPS spectrum with template spectra of real stars with 
well known parameters via a $\chi^2$ minimization criterion. 
The derived stellar parameters come from a weighted mean of the 
parameters of the 50 reference stars (10 per each spectral region) that 
independently match the target spectrum most closely. The results are 
$T_{\rm eff}$$=3460\pm60$\,K, $\log{g}$$=0.6\pm0.2$, 
[Fe/H]$=-0.05\pm0.10$, and v$\sin{i}$$=6.2\pm3.0$\,km~s$^{-1}$.
From the $\log{g}$ vs. $T_{\rm eff}$ relationship by \citet{houdashelt00} 
for solar metallicity giants, we estimate $\log{g(3500\,K)}\sim0.4$, 
in fair agreement with the value derived from the $\chi^2$ 
minimization. 

\subsection{Lithium abundance}
As pointed out by \citet{gratton89}, a high value ($\sim$500\,m\AA)
of the $\lambda$6707.8\,\AA\ Li\,{\sc i} line equivalent width implies 
that the resonance doublet should be strongly saturated and that 
deviations from LTE conditions may be important. This requires the 
use of other lithium lines. The spectral range covered by HARPS 
also allows us to investigate the $\lambda$6103.6\,\AA\ Li\,{\sc i} 
line.

\begin{figure}[ht] 
\vskip 9.5cm
\includegraphics{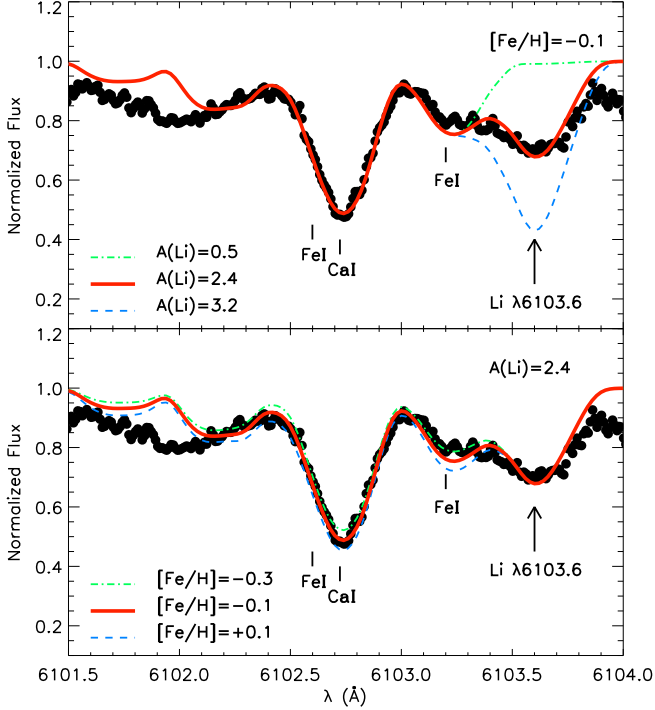}
 \caption{Portion of HARPS spectrum in the interval around the 
 lithium $\lambda$6103.6\,\AA\  absorption line. The top panel shows 
 the synthetic spectra for three values of $A$(Li) at 
 fixed metallicity, while the lower one shows the synthetic spectra 
 for three values of metallicity at fixed $A$(Li). The best fit
 is for a sligthly subsolar metallicity and $A$(Li)=2.4.}		
 \label{NLi_03}
\end{figure}

The Li abundance $A$(Li) was derived by spectral synthesis of the 
\ion{Li}{i} 6103.6\,\AA\  line using the MOOG code \citep{sneden73}, 
which assumes LTE conditions, along with the GAIA model 
atmospheres (Brott \& Hauschildt 2010, priv. comm.). 
The hyperfine line structure has been considered following 
the guidelines by \citet{wahlgren05} to compute the atomic parameters.
The atomic and molecular line lists were taken from the 
VALD\footnote{http://vald.astro.univie.ac.at/$\sim$vald/php/vald.php.} 
database.
The physical parameters derived from the $\chi^2$ minimization explained 
above were used as starting values for the synthesis. We let 
metallicity, $v\sin{i}$, $A$(Li), and microturbulence velocity $\xi$ vary,
while keeping $T_{\rm eff}$ and $\log{g}$ fixed. 
The low temperature of the star makes the spectral synthesis rather 
difficult, because of the uncertainty on continuum determination caused 
by several overlapping molecular lines. The continuum was determined 
by a spline fit made after visually selecting regions free of absorption 
lines. We estimated that the uncertainty on the continuum determination,
which is the main source of error on $A$(Li), is less than 5\%. The results are 
shown in Fig.~\ref{NLi_03}. From our best fit, we obtained values of 
$A$(Li)$=2.4\pm0.2$, [Fe/H]$=-0.1\pm0.2$, and $v\sin{i}=9\pm2$\,km~s$^{-1}$ 
for a microturbulence velocity $\xi=0.8\pm0.2$\,km\,s$^{-1}$. 
The [Fe/H] and $v\sin{i}$ values are in good agreement with those from 
the $\chi^2$ minimization, so average values were computed. The final 
adopted physical parameters are reported in Table~\ref{physpar}.

An attempt at spectral synthesis of the lithium line at 6707.8\,\AA\   
with MOOG (using the \citealt[][]{reddy02} line list implemented within 
the VALD database) resulted in large residuals (c.f. Fig.~\ref{NLi_07}), 
mainly owing to non-LTE conditions, such as the effects of over-ionization, 
and line asymmetry due to the convective motions, which mainly 
influence the $\lambda$6707.8\,\AA\ Li\,{\sc i} line formation in 
very cool stars (\citealt{carlssonetal1994}). Since the EW of this line 
could be measured with high accuracy, we could estimate the non-LTE Li 
abundance through extrapolating the curves of growth by \citet{lind09} 
to a temperature of $\sim$3500\,K, which yield $A{\rm (Li)}\sim$2.5. 

We thus conclude that a reliable value for the lithium abundance 
of \iras\ is $A$(Li)$=$$2.4\pm0.2$. As a result, the low temperature and high 
lithium abundance of the star make \iras\ the coolest lithium-rich giant 
known to date. As shown below, our determinations of temperature and 
gravity place the star close to the tip of the red giant branch (RGB) 
on the HR diagram.

\section{Discussion and conclusions}
\begin{figure}[ht] 
\vskip 10.0cm
\includegraphics{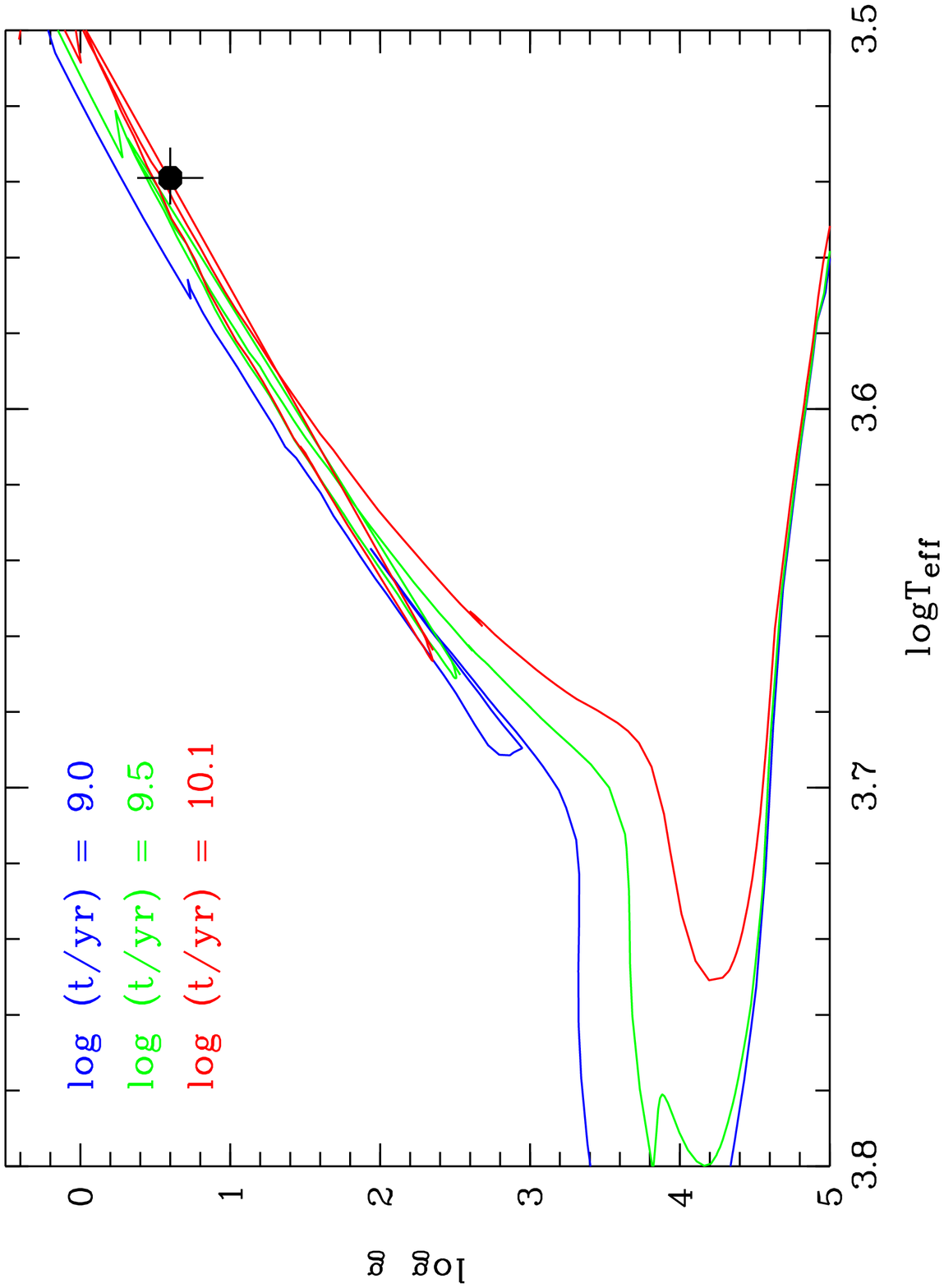}
\includegraphics{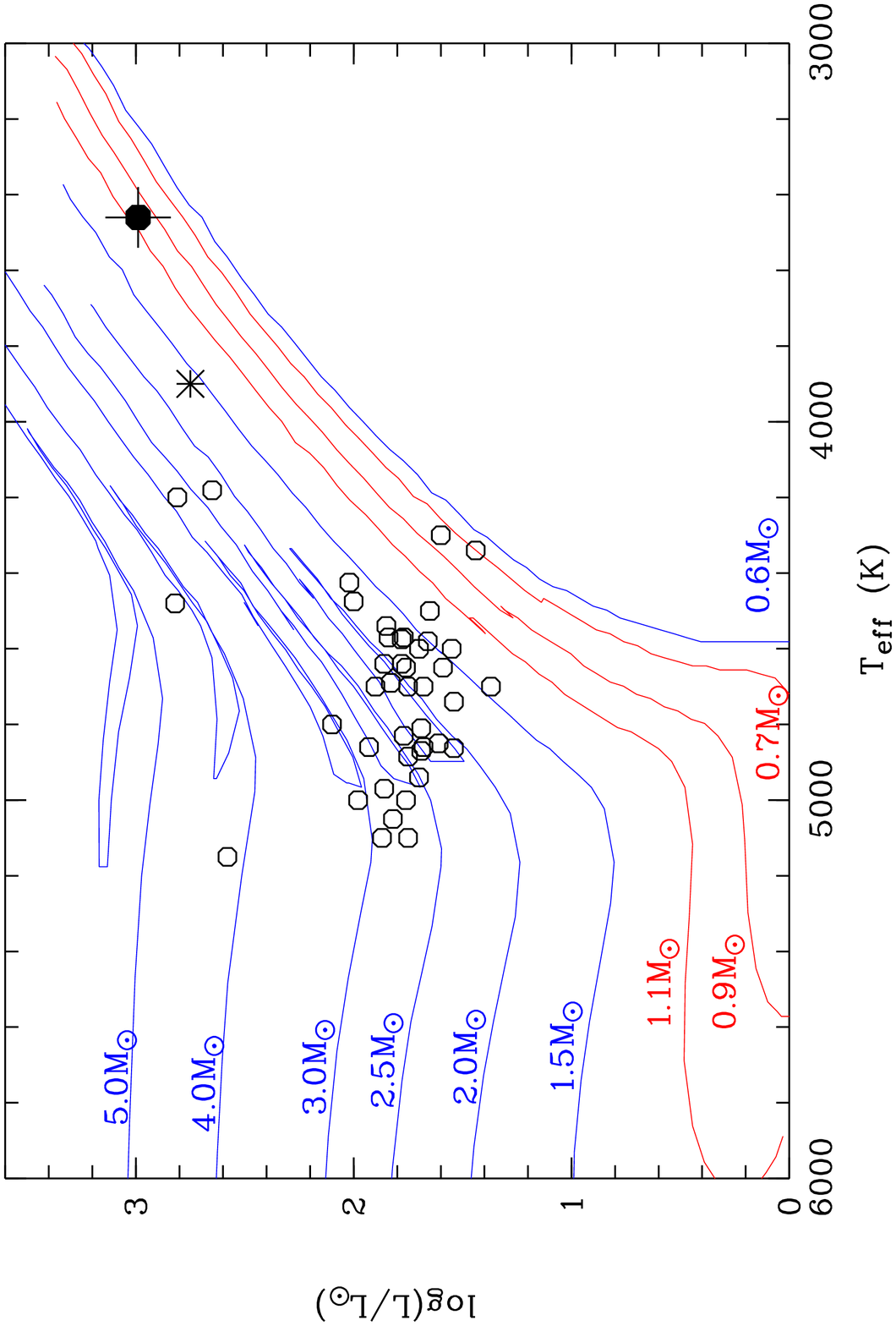}
 \caption{{\em Upper panel}. The $\log{g}$ versus $\log{T_{\rm eff}}$ diagram. The 
  \citet{girardi00} isochrones for three different ages are overplotted.
  The black dot represents the position of \iras.
  {\em Lower panel}. HR diagram of several Li-rich giants. The evolutionary 
  tracks by \citet{girardi00} for several masses as labelled are represented with 
  continuous lines. The lithium-rich K-type giants \citep{kumar11} 
  are over-plotted with open circles. The asterisk represents 
  HD~39853, while the position of \iras\  is indicated by the 
  big black dot.}
 \label{HRD}
\end{figure}

 The stellar parameters of \iras\ indicate that the star is a lithium-rich M-giant. 
 To investigate its evolutionary status in more detail, we used 
 the tracks and isochrones by \citet{girardi00} for solar metallicity stars. 
 A comparison of the position of \iras\ in the $\log{g}$ vs. $\log{T_{\rm eff}}$  
 diagram (Fig.~\ref{HRD}) showed that the physical parameters are consistent 
 with an age $\sim$10\,Gyr. Reasonable values for the stellar luminosity and 
 mass at that age, which are consistent with $\log{T_{\rm eff}}$, are 
 $\log{(L/L_{\odot})}=$2.99$\pm$0.15 and $M/M_{\odot}=$1.0$\pm$0.2, 
 respectively. The position of \iras\ on the HR diagram is shown in Fig.~\ref{HRD}, 
 along with several lithium-rich stars, as compiled by \citet{kumar11}. 
 Most of the known lithium-rich giants are concentrated within a luminosity 
 range of 1.3$< \log{(L/L_{\odot})}<$2, with only a minority having 
 $\log{(L/L_{\odot})}>$2.2. Among these luminous K-type objects, HD~39853 
 is the coolest one; nevertheless, \iras\ is about 450\,K cooler, placing it 
 among the least massive and most luminous lithium-rich giants known so far. 
 Being close to the tip of the RGB, one should then consider three 
 possibilities for the evolutionary stage of the star:
 {\it i) }~giant branch ascent, i.e. H-shell burning phase;
 {\it ii)}~AGB phase; or
 {\it iii)}~post He-core flash.
 Unfortunately, the CNO abundances and the $^{12}$C/$^{13}$C carbon isotopic 
 ratios, tracers of the degree of mixing that provide further constraints on 
 the evolutionary status, cannot be derived because the spectral range of 
 HARPS does not include the appropriate wavelength range. 
   
 Assuming a distance $d_{\rm Cha\,II}=178$\,pc, \citet{spezzi08} have
 derived a luminosity $\log{(L/L_{\odot})}\approx$1. By comparing 
 it with the luminosity estimated above, we conclude that 
 \iras\ should be located at $d\approx$1.75\,kpc from the Sun.
 By adopting this distance and using the HARPS radial velocity and the 
 proper motion components (cf. Sec.~\ref{Sec:spectra}), the resulting 
 spatial velocity components of the star in a left-handed coordinate 
 system are ($U, V, W$)$=$($+30, -121, +79)$~km\,s$^{-1}$, where 
 $U$, $V$, and $W$ are directed towards the Galactic anti-centre, 
 the Galactic rotation direction, and the North Galactic Pole,
 respectively. According to the criteria of 
 \citet{oort26}\footnote{ $ | W + 10 | \ge $30\,km~s$^{-1}$ 
 and/or ($U^2 + V^2$)$^{1/2} \ge $65\,km~s$^{-1}$.}, 
 \iras\ can thus be considered as a high-velocity star.
 The Galactic latitude, distance, and kinematics of the star imply that 
 it most likely belongs to the old thin-disk population, consistent 
 with its almost solar metallicity. But what is the cause of its high 
 lithium content?

 Three possible scenarions can be considered to explain the high lithium 
 abundance observed in \iras:
 {\it i)} the star somehow preserved the original lithium in its atmosphere, 
 {\it ii)} lithium has been regenerated in later evolutionary stages by the
  Cameron--Fowler mechanism\footnote{Conversion of $^3$He to $^7$Li by
 $\alpha$-capture with $^7$Be as a radioactive intermediary.} \citep{cameron71},
 {\it iii)} lithium was enhanced by the engulfment of a brown dwarf or 
 planetary companion.

 As discussed in \citet{fekel93}, the scenario of the preservation of initial 
 lithium in red giants is very unlikely. They also note that high lithium 
 abundance was not revealed in a sample of 200 F-type stars with masses 
 between 1 and 2\,$M_{\odot}$, just evolved off the main sequence. 
 As they point out, if anything, lithium preservation might eventually 
 work in stars of the early-F and A types, that are more massive than \iras. 
 As a solar mass star, the initial lithium was most likely 
 burned during the pre-main sequence phase. 

\begin{figure}[ht] 
\vskip 6.0cm
\includegraphics{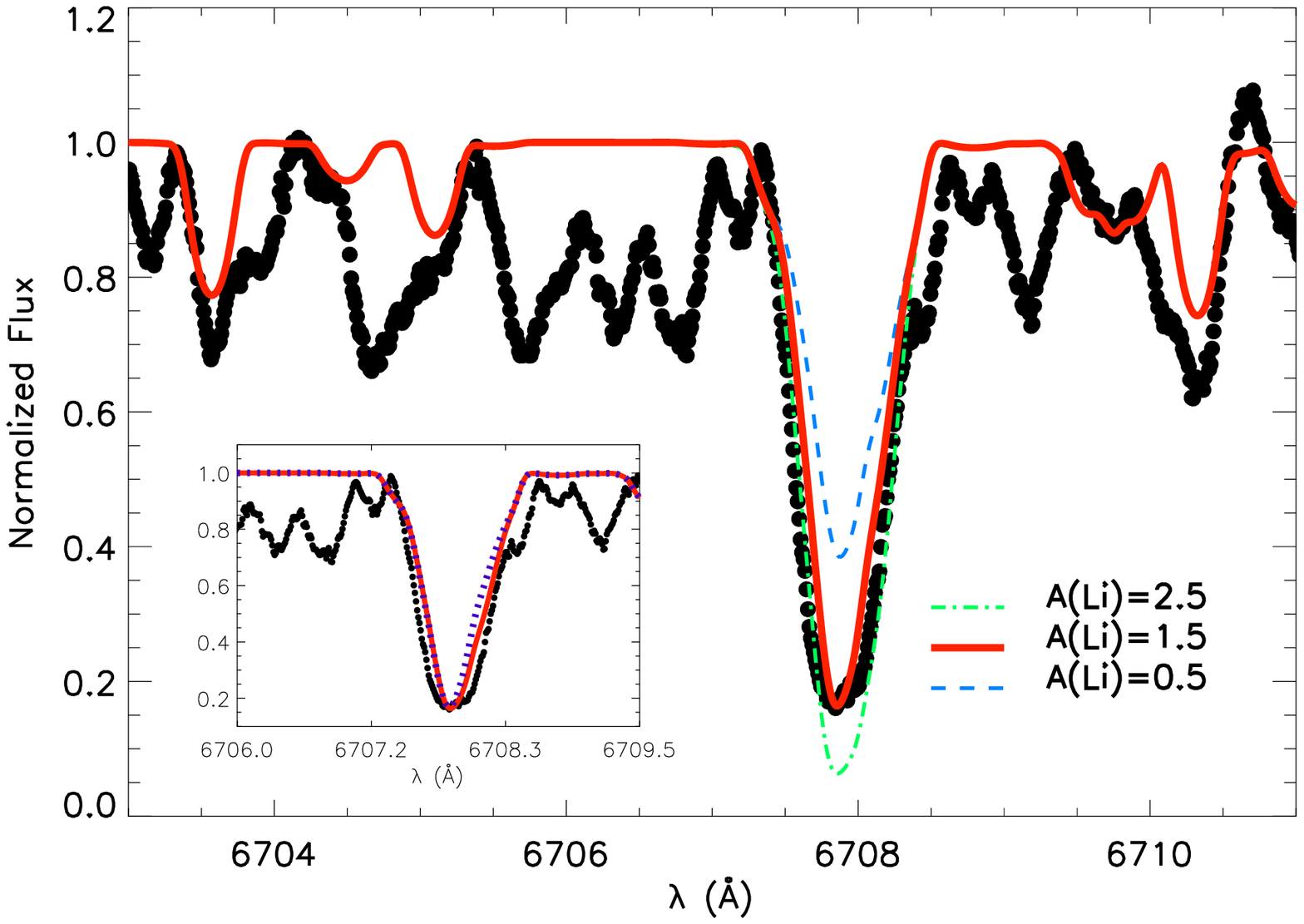}
 \caption{Comparison of the observed lithium $\lambda$6707.8\,\AA\  absorption line
  with synthetic MOOG spectra for three values of $A$(Li).  
  The $^6$Li/$^7$Li isotopic ratio has been also included as a free 
  parameter for the fit. The inset shows the comparison of the results 
  with and without inclusion of the $^6$Li/$^7$Li ratio as a free 
  parameter (continuous and dotted lines, respectively).} 
 \label{NLi_07}
\end{figure}

 Depending on the stellar mass and luminosity, the production of lithium 
 via the Cameron--Fowler mechanism supossedly occurs both at
 the RGB luminosity function bump \citep{charbonnel00} and during the 
 He-core flash \citep{kumar11}. It is possible that lithium was 
 synthesized in \iras\ by the Comeron--Fowler mechanism during the 
 luminosity bump, but most likely it had already been destroyed as the star 
 evolved up to the RGB. Although the \citet{sackmann99} model predicts that 
 a high lithium abundance may be preserved all the way up the RGB, their 
 parameterization requires very high mixing rates. It was assumed that 
 rotation could provide such rates, but \citet{palacios06} find that a 
 self-consistent model of rotational mixing cannot generate enough 
 circulation to account for the mechanism working efficiently.
 It has been suggested \citep{delareza96} that the IR-excess observed 
 in some lithium-rich giants may be due to a dust shell possibly also
 generated via the Cameron-Fowler mechanism, but for \iras\ there 
 is no evidence of IR excess \citep[][]{larson98, alcala08}.
 On the other hand, it is unlikely that, assuming an AGB status for the 
 star, lithium has been synthesized during the He-flash phase, because
 such a process should work for stars within a narrow mass range around 
 2$M_{\odot}$ \citep{kumar11}. 
 A pure Cameron-Fowler mechanism should  only provide $^7$Li, with no 
 $^6$Li. 
 An attempt at a simultaneous fit of the resonance and subordinate 
 lithium line by including the $^6$Li/$^7$Li isotopic ratio as a free 
 parameter improves the fit of the $\lambda$6707.8 line
 (see Fig.~\ref{NLi_07}), but there is  no way to get a good fit 
 for values of $^6$Li/$^7$Li less than 0.11.

 In the brown dwarf/planet engulfment scenario, the accreted matter 
 would also result in a simultaneous enhancement of $^6$Li, $^7$Li,  
 and Be. Unfortunately, our HARPS spectrum does not achieve the wavelength 
 range to investigate beryllium, but the suggestion that the $^6$Li/$^7$Li  
 isotopic ratio may be as high as 0.11 would support the accretion 
 scenario. Also, the large radius of \iras\ in comparison with most of 
 the known lithum-rich giants makes the planet engulfment scenario 
 plausible, because planet ingestion would be more likely to occur when 
 a star evolves more in the RGB and achieves a larger radius.

 Finally, in discussing the case of \iras, we have to consider that its 
 projected rotational velocity, $v\sin{i}\sim$8\,km s$^{-1}$, is rather 
 high in comparison with other lithium-rich giants. Several authors have 
 argued that accretion of a planet may explain both lithium enhancement 
 and rapid rotation in giants
 \citep[][and references therein]{denissenkov04,calsberg09,calsberg10}.
 Given its physical properties, we cannot rule out that such a 
 process has led to the lithium enhancement and rapid rotation 
 in \iras. Some rapidly rotating giants can even become magnetically active 
 and may be detected in X-rays \citep[][]{guillout09}, however \iras\ was not
 detected in a ROSAT pointed observation \citep{alcala00}.

 With the observational data available so far, we cannot be conclusive
 about what process dominates the lithium enhancement in \iras.
 Perhaps several mechanisms working at different times along the RGB 
 phase of this star have contributed to its lithium enrichment.
    
\acknowledgements{We thank the referee, Dr. P. Bonifacio, for his 
useful comments and suggestions and for information on the lithium 
hyperfine line structure. K.B. acknowledges financial support from the 
INAF Postdoctoral fellowship programme. We also thank V. Andretta, 
L. Belluzzi, and V. D'Orazi for discussions of hyperfine line structure.}

\bibliographystyle{/home/jmae/aa-package/bibtex/aa}

\Online

{\bf To be published in electronic form only}

\begin{appendix}

\section{The ROTFIT procedure} \label{appendix1}

\begin{figure}[ht] 
\vskip 9.0cm
\includegraphics{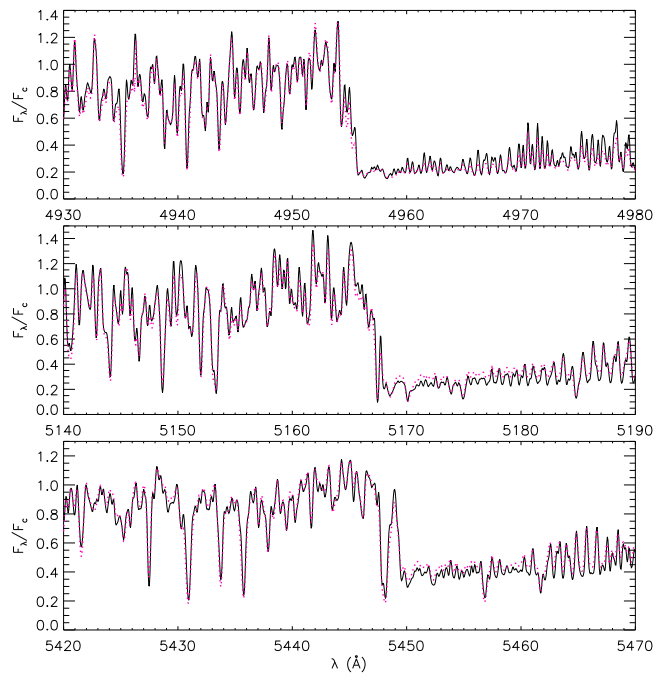}
 \caption{Portions of the HARPS spectrum of \iras\ around three TiO bands. 
 The M5\,III best template (\object{HD\,175865}) is overplotted with 
 dotted red lines.}
 \label{harps_TiO}
\end{figure}

In order to derive the astrophysical parameters of \iras, we used 
a $\chi^2$ minimization  procedure developed in the 
IDL\footnote{IDL (Interactive Data Language) is a registered trademark 
of IT Visual Information Solutions.}
environment. This procedure, which is a variation of the ROTFIT 
code described by \cite{frasca03}, provides us with the best match of 
the observed spectrum with a grid of high-resolution spectra of real stars 
with well determined parameters \citep[PASTEL catalogue,][]{soubiran10} 
retrieved from the ELODIE archive \citep{moultaka01}, which is one of the 
largest datasets of high-resolution spectra, spanning the wavelength 
range 4000--6800\,\AA. Because the spectra are  severely affected by molecular 
bands, it was difficult to perform a safe and homogeneous normalization 
to the local continuum for both \iras\ and the ELODIE templates. 
We thus preferred to analyse selected spectral regions, mainly around 
the strongest TiO bands ($\lambda\lambda$4950, 5166, 5450, 6158, 6650, 6680\,\AA), 
normalizing the spectrum with respect to the average stellar flux in a 
window of $\sim 10$\,\AA\  blueward of the band-head (see Fig.~\ref{harps_TiO}). 
This allowed us to fully exploit the high sensitivity to both temperature 
and gravity of the molecular bands \citep[see, e.g.,][]{neff95}. 
The \ion{Na}{i}\,D lines, which are very sensitive diagnostics of temperature 
and gravity, were selected as well (Fig.~\ref{harps_spec}). 
At the end of this procedure, for each of the five spectral regions 
investigated, ten templates that best matched the HARPS spectrum of \iras, 
based on $\chi^2$ minimization, were chosen independently, and their 
known astrophysical parameters were used to compute weighted average values.
Such average values were adopted as astrophysical parameters for \iras. 
The results are $T_{\rm eff}=3460\pm60$\,K, $\log{g}=0.6\pm0.2$\,dex, 
[Fe/H]\,$=-0.05\pm0.10$\,dex, and v$\sin{i}=6.2\pm3.0$\,km\,s$^{-1}$.
The errors include the 1$\sigma$ standard deviation on the average 
and the typical errors of the PASTEL astrophysical parameters added 
in quadrature. The latter are $\pm$50\,K, $\pm$0.1\,dex, $\pm$0.1\,dex,
and $\pm$0.5\,km~s$^{-1}$ for $T_{\rm eff}$, $\log{g}$, [Fe/H], and
v$\sin{i}$, respectively. 

\end{appendix}

\end{document}